\documentclass[letterpaper]{article} 
\usepackage{aaai2026}  
\usepackage{times}  
\usepackage{helvet}  
\usepackage{courier}  
\usepackage[hyphens]{url}  
\usepackage{graphicx} 
\usepackage{natbib}  
\usepackage{caption} 
\usepackage{fancyhdr}
\fancypagestyle{firstpage}{
  \fancyhf{}
  \fancyhead[R]{\scriptsize Accepted by AAAI 2026}
  
}
\usepackage{algorithm}
\usepackage{algorithmic}
\usepackage{enumitem}
\usepackage{tabularx}
\usepackage{dsfont}
\usepackage{subcaption}
\usepackage{multirow}
\usepackage{amsmath}
\usepackage{pgfplots}
\usepackage{tabularx}
\usepackage{booktabs}
\usepackage[T1]{fontenc}
\usepackage{newfloat}
\usepackage{listings}
\usepackage{xcolor}
\usepackage{times}
\usepackage{helvet}
\usepackage{courier}
\usepackage{xcolor}
\usepackage{newfloat}
\usepackage{listings}
\DeclareCaptionStyle{ruled}{labelfont=normalfont,labelsep=colon,strut=off} 
\floatstyle{ruled}
\newfloat{listing}{tb}{lst}{}
\floatname{listing}{Listing}
\title{Beyond Static: Related Questions Retrieval Through Conversations in
Community Question Answering}

\author{
    Xiao Ao\textsuperscript{\rm 1},
    Jie Zou\textsuperscript{\rm 1}\thanks{Jie Zou and Weikang Guo are corresponding authors.},
    Yibiao Wei\textsuperscript{\rm 1},
    Peng Wang\textsuperscript{\rm 1},
    Weikang Guo\textsuperscript{\rm 2 $*$}
}
\affiliations{
    \textsuperscript{\rm 1}School of Computer Science and Engineering, University of Electronic Science and Technology of China\\
    \textsuperscript{\rm 2}School of Management Science and Engineering, Southwestern University of Finance and Economics\\

    202422080439@std.uestc.edu.cn, jie.zou@uestc.edu.cn, 202411081540@std.uestc.edu.cn, p.wang6@hotmail.com, guowk@swufe.edu.cn
}

\usepackage{bibentry}

\begin{document}

\maketitle
\thispagestyle{firstpage}
\begin{abstract}
In community question answering (cQA) platforms like Stack Overflow, related question retrieval is recognized as a fundamental task that allows users to retrieve related questions to answer user queries automatically. Although many traditional approaches have been proposed for investigating this research field, they mostly rely on static approaches and neglect the interaction property. We argue that the conversational way can well distinguish the fine-grained representations of questions and has great potential to improve the performance of question retrieval. In this paper, we propose a related question retrieval model through conversations, called TeCQR, to locate related questions in cQA. Specifically, we build conversations by utilizing tag-enhanced clarifying questions (CQs). In addition, we design a noise tolerance model that evaluates the semantic similarity between questions and tags, enabling the model to effectively handle noisy feedback. Moreover, the tag-enhanced two-stage offline training is proposed to fully exploit the mutual relationships among user queries, questions, and tags to learn their fine-grained representations. 
Based on the learned representations and contextual conversations, TeCQR incorporates conversational feedback by learning to ask tag-enhanced clarifying questions to retrieve related questions more effectively. Experimental results demonstrate that our model significantly outperforms state-of-the-art baselines.

\end{abstract}
\begin{links}
    \link{Code}{https://github.com/AIT55/TeCQR}
\end{links}
\section{Introduction}
Community question answering (cQA) platforms have become vital resources for knowledge sharing and acquisition \cite{stackoverflow1, stackoverflow2}. On platforms such as Stack Overflow (SO), users often prefer to retrieve answers from resolved questions rather than post new queries and wait for replies. Therefore, effective question retrieval plays a critical role in enhancing the utility of cQA platforms.

\begin{figure}[tbp]
    \centering
    \includegraphics[width=1.0\linewidth]{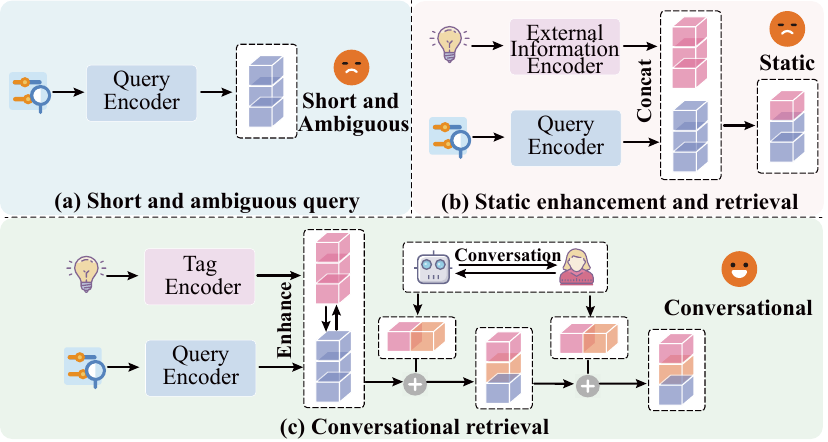}
    \caption{(a) Short and ambiguous query hinder intent understanding. (b) Static enhancement and one-off interaction may introduce unconfirmed information and fail with initially unclear user intent. (c) We iteratively incorporate user-confirmed information through tag-based clarifying questions to elicit more accurate user intent for retrieval.}
    \label{fig:intro}
\end{figure}

For question retrieval in cQA, a wide range of methods have been proposed \cite{cnnMulti-layer, RCNN, answerquality, Bert1, tagco, questionstag}. Most of these approaches \cite{RCNNStr, useGloVe2, Lit, Litword2vec, Dmso, rmso, useglove1, useglove3, Bilstmcnn, AsymmetricRanking, usesenbert, bertcross, bertandkg} embed both the query and candidate questions as vectors, then measure their similarity in explicit or latent semantic space. However, queries in question retrieval are often short and ambiguous (Figure~\ref{fig:intro}(a)), making accurate retrieval particularly challenging.

To alleviate the above challenge, several studies have attempted to enhance query representations by incorporating external information, such as tags associated with the query \cite{tagco, tagenh}, knowledge graphs \cite{bertandkg}, answerer authority \cite{AsymmetricRanking}, answer content \cite{multiatten, attentionjz}, and other contextual signals \cite{qnogood1, qnogood2, QUERYEXP2, shen1, shen2}. Although these methods have shown effectiveness in enriching query information, they typically rely on the static information enhancement and one-off query interaction to perform static retrieval (Figure~\ref{fig:intro}(b)), yet still suffer from two limitations. First, external information that has not been confirmed by the user may be unreliable, or even misaligned with the user’s actual intent (e.g., when a user intends to understand the usage of \textit{pop()} function in a dictionary but does not explicitly specify \textit{dict}, the system may incorrectly associate \textit{list} as external information, leading to irrelevant retrieval outcomes.). Such mismatches can lead the retrieval results to deviate from the user’s actual intent. Second, users often struggle to articulate their intent clearly when they first submit a query, especially when they are unfamiliar with relevant technical terms \cite{userstud1, userstud2}. Inspired by research on clarifying questions \cite{cq1, cq2, cq3, cq4, learning1, askma, askcq} in related information retrieval domains, conversational retrieval with clarifying questions offers a natural solution to these issues. By iteratively introducing user-confirmed information by clarifying questions, this approach helps guide users toward expressing a more complete and accurate intent for related question retrieval.

Building upon the idea of clarifying questions, and aiming to address both the limitations of external information and the ambiguity of users’ initial queries, we propose a novel \textbf{T}ag-\textbf{e}nhanced \textbf{C}onversational \textbf{Q}uestion \textbf{R}etrieval model, called \textbf{TeCQR}(Figure~\ref{fig:intro}(c)), to retrieve related questions effectively through conversations. To begin with, we utilize a pre-trained language model to initialize separate embeddings for queries, questions, and tags. Then, we construct a \textit{tag-enhanced two-stage offline training} framework to enhance the representations for more effective conversational retrieval, which consists of \textit{query–question training via conversation simulation} and \textit{tag–question training via contrastive learning}. Additionally, we introduce a \textit{noise tolerance} model to mitigate the impact of noisy user feedback. Building on this, the \textit{tag-enhanced conversational retrieval} phase selects an optimal sequence of tags to form clarifying questions and iteratively augments the query via user-confirmed tag information. After the conversations built upon clarifying questions, we utilize a soft-matching method to iteratively integrate user feedback to locate related questions effectively. 

Our main contributions are summarized as follows:
\begin{itemize}[left=0pt]
\item 
We formalize the paradigm of Conversational Question Retrieval (CQR) and propose a novel CQR model, named TeCQR, which enables users to clarify their intent through tag-enhanced conversation.
\item We design a tag-enhanced two-stage offline training to enhance the representations of queries, questions, and tags, making them more effective for conversational retrieval. 
\item We introduce a noise tolerance modeling to effectively handle noisy user feedback.
\item Extensive experimental results demonstrate that TeCQR significantly outperforms existing state-of-the-art baselines.
\end{itemize}

\begin{figure*}[t] 
    \centering
    \includegraphics[width=0.9\textwidth]{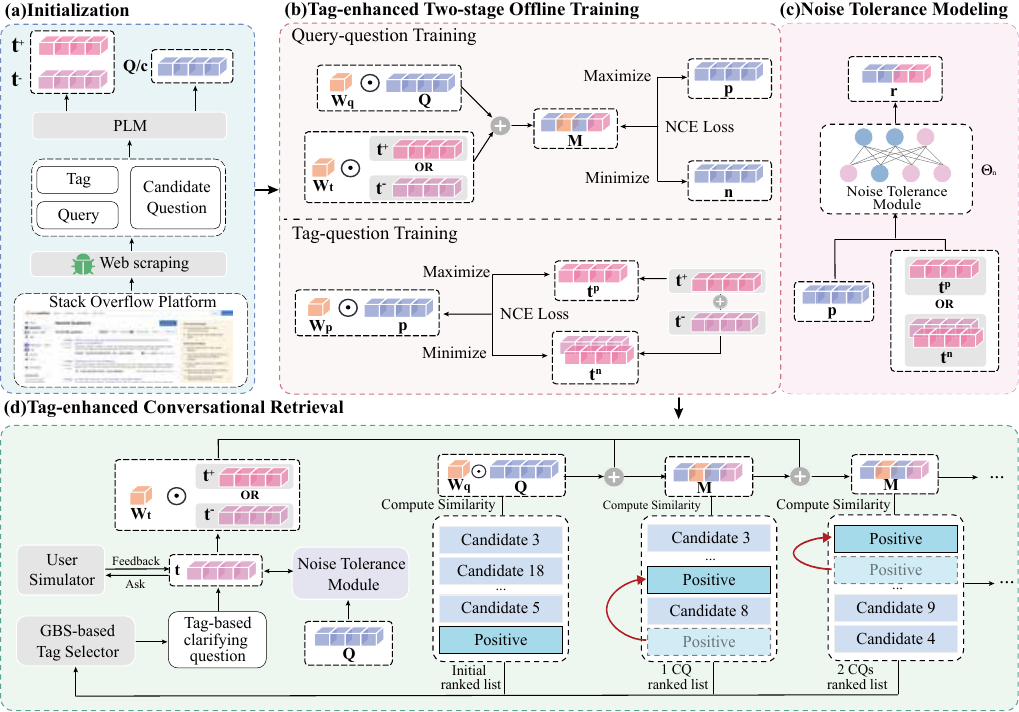} 
    \caption{Overview of TeCQR.}
    \label{fig:model structure}
\end{figure*}

\section{Related Work}
Question retrieval is the task of retrieving related questions in response to a user-submitted query on cQA platforms. Early methods like BM25 \cite{bm25} relied on exact term matching, lacking semantic understanding. Later work \cite{gaosi, userstud1} introduced classical machine learning to incorporate semantic features.

Despite these advancements, classical machine learning models still struggled with capturing nuanced semantics, prompting a shift toward pre-trained embeddings and deep learning approaches for question retrieval \cite{useglove1, useGloVe2, useglove3, cnnMulti-layer, bertandkg, Bert1, usesenbert, bertcross, Lit, Litword2vec}. BiLSTM-based models constructed semantic interaction modules to model the relationships between queries and candidate questions \cite{Dmso, rmso, ASIM, Bilstmcnn}. Similarly, RCNN \cite{RCNN, RCNNStr} utilized gated convolutions to encode questions, distilling key information from potentially noisy text. HCA \cite{wordandsentence} further enriched representations by combining sentence and word-level cues. However, these deep models still suffer from short and ambiguous queries that lack sufficient contextual cues for accurate retrieval.

To alleviate this limitation, prior work leverage external information on cQA platforms. Query expansion was explored to handle short and ambiguous queries \cite{qnogood1, qnogood2, QUERYEXP2}, but it often introduced irrelevant or non-technical terms, increasing retrieval complexity. To address this, tags were widely applied to refine query representations. For example, \citet{tagco} learned tag representations with node2vec on a tag co‑occurrence graph, \citet{questionstag} derived distributed tag embeddings by applying DeepWalk \cite{deepwalk} to a tag graph, and \citet{tagenh} embed questions by concatenating the tag sequence with the question text. However, these methods were developed for duplicate question retrieval, where questions in the database are already labeled with tags. This setting differs from the real-world scenario we focus on, where users submit tagless queries to cQA platforms. Additionally, \citet{chatbot} conducted an initial user study with human volunteers, demonstrating the potential of asking CQs to improve retrieval accuracy. Beyond tags, other studies have leveraged external signals, including knowledge graphs \cite{bertandkg, moekg}, the authority of answer providers \cite{AsymmetricRanking, answerquality}, and answer content \cite{multiatten, attentionjz}.

However, the above methods treat external information and questions as separate components, relying on static information enhancement and one-off query interaction. In contrast, we innovatively introduce a tag-enhanced conversational fusion mechanism that dynamically incorporates external information during the retrieval process. By guiding user interaction in an incremental manner, our approach enables a more accurate understanding of query intent and enhances retrieval performance.

\section{Problem Formalization}
Given an initial query \( Q \) submitted by a user, the goal of question retrieval is to identify a set of related questions \( P = \{p_1, p_2, \dots, p_k\} \) from a candidate pool \( C = \{c_1, c_2, \dots, c_n\} \), where \( P \subseteq C \). The ground-truth set \( P \) is annotated by real users on the Stack Overflow platform, while the remaining candidates \( N = C \setminus P = \{n_1, n_2, \dots, n_j\} \) are treated as negative examples. The typical retrieval objective can be formally expressed as:
\begin{equation}
Q \rightarrow \{ \text{ranked}(c_1, c_2, \dots, c_n) \} \rightarrow P.
\label{eqqp}
\end{equation}

In our work, the retrieval process is enhanced by introducing clarifying questions to interact with the user. Specifically, the system asks a clarifying question \( q^t \) based on a specific tag \( t \) during the retrieval session. The user responds with feedback \( f \), forming a clarifying question \& feedback pair, which is denoted as \( (q^t, f) \). Each retrieval session contains multiple rounds of conversation. Thus, a sequence of clarifying questions is asked to the user, and a sequence of associated feedback is collected for these questions. Suppose \( l \) rounds of clarifying questions have been asked, the sequence of clarifying question \& feedback pairs can be represented as: $\{(q^t_1, f_1), (q^t_2, f_2), \dots, (q^t_l, f_l)\}$. The sequence of actions in conversational question retrieval can be represented as follows:
\begin{equation}
Q \rightarrow \{ \text{ranked}(c_1, c_2, \dots, c_n) \}, \{(q^t_i, f_i)\}_{i=1}^l \rightarrow P.
\label{eq1}
\end{equation}

\section{Methodology}
In this section, we introduce our proposed TeCQR model. An overview of TeCQR is shown in Figure \ref{fig:model structure}. The TeCQR model consists of four modules: (a) initialization using pre-trained language models; (b) tag-enhanced two-stage offline training, including query-question training via conversation simulation and tag-question training via contrastive learning to enhance these embeddings; (c) noise tolerance model, which effectively handle noisy user feedback; (d) tag-enhanced conversational retrieval, which iteratively refines user intent through tag-based clarifying questions to improve retrieval performance. The dataset construction process is described in \textit{Section of Dataset}.
\subsection{\textbf{Initialization}}
To place query, candidate question, and tag in a unified semantic space, we share a single encoder, the pre-trained language model \texttt{All-MiniLM}~\cite{allminilm}, to encode them into representations $\mathbf{Q}$, $\mathbf{c}$, and $\mathbf{t}$, respectively, as illustrated in Figure~\ref{fig:model structure}(a). Since queries and candidate questions share the same format, both are encoded using All-MiniLM:
\begin{equation}
\mathbf{Q} = \mathbf{F}_{\text{All-MiniLM}}(i_Q), \quad \mathbf{c} = \mathbf{F}_{\text{All-MiniLM}}(i_c),
\end{equation}
where \( i_Q \) and \( i_c \) are the corresponding textual inputs of query or candidate questions. 

Modeling negative feedback is particularly challenging, as relevant items often share similar characteristics, while the reasons for irrelevance can vary significantly. To address this, we learn a separate embedding for negative feedback, which serves as a strong supervisory signal and plays a crucial role in guiding the retrieval process. Specifically, we encode the tag text $i_t$ as:
\begin{equation}
\mathbf{t}^+ = \mathbf{F}_{\text{All-MiniLM}}(i_t), \quad \mathbf{t}^- = -\mathbf{t}^+, 
\end{equation}
where positive tag representation $\mathbf{t}^+$ corresponds to a tag that is semantically relevant to the query (i.e., receives positive feedback), while the negative tag representation $\mathbf{t}^-$ is constructed by negating $\mathbf{t}^+$, aiming to capture the semantics of negative feedback.
\subsection{\textbf{Tag-enhanced Two-stage Offline Training}} 
\textbf{Query-question Training via Conversation Simulation.} 
In query–question training, we aim to capture the relationship between a query and its related questions. To achieve this, we first enhance the query representation by simulating the process of asking clarifying questions and receiving user feedback. Specifically, at each round of the conversation, we pose a tag-based clarifying question and receive user feedback on the corresponding tag. To incorporate this feedback into the query representation, we construct a mixture query representation $\mathbf{m}$ as follows:
\begin{equation}
\mathbf{m} = W_Q \cdot \mathbf{Q} + W_t \cdot \mathbf{t}^{\pm}, \label{eq:mixture_query}
\end{equation}
where $\mathbf{Q}$ denotes the original query embedding, and $\mathbf{t}^{\pm}$ represents either the positive tag embedding $\mathbf{t}^{+}$ or the negative tag embedding $\mathbf{t}^{-}$, depending on the user feedback. In conversational retrieval, the query and tag contribute unequally to identifying the target question. We introduce learnable weights \( W_Q \) and \( W_t \) to adaptively balance their contributions by capturing their mutual relationship.

After enhancing the query representation with user feedback from clarifying questions, we use the refined query for retrieval. Specifically, we adopt a Noise Contrastive Estimation (NCE) loss~\cite{nce} as follows:
{\small
\begin{equation}
\begin{split}
\mathcal{L}_{\text{QQ}} = &- \frac{1}{|M|} \sum_{\mathbf{m} \in M} 
\Bigg[ 
\log \sigma(\mathbf{m}^\top \mathbf{p}) \\
&+  \frac{1}{|N|} \sum_{\mathbf{n} \in N} \log \left( 1 - \sigma(\mathbf{m}^\top \mathbf{n}) \right)
\Bigg],
\end{split}
\end{equation}}where \( \mathbf{p} \) denotes the embedding of the positive (target) question, and \( \mathbf{n} \in N \) denotes the representation of a negative question sample. During training, each query is paired with its corresponding positive question(s) as labeled by the Stack Overflow platform. If multiple positive questions exist for a given query, one is randomly selected in each training round. Negative samples are randomly drawn from the entire question pool. \( \sigma(x) \) is the sigmoid function. \( |M| \) and \( |N| \) denote the total number of mixture queries and the number of negative samples per query, respectively. \\
\textbf{Tag-question Training via Contrastive Learning.} 
Since each tag captures a specific aspect of a question, a combination of relevant tags can serve as a rough semantic approximation of the question. Therefore, in the embedding space, a question representation should be closer to its associated tags and farther from unrelated ones. To enforce this, we introduce the tag-question training, which models the mutual relationship between questions and tags. 

To adjust the relative contribution of questions and tags during alignment, we also employ an adaptive parameter on the question representation. Specifically, the adjusted question \( \mathbf{p'}\) representation is computed as:
\begin{equation}
\mathbf{p'} = W_{p} \cdot \mathbf{p},
\label{eq:qu}
\end{equation}
where \( \mathbf{p} \) denotes the original question embedding, and \( W_{p} \) is a learnable parameter that controls the contribution of the question in the alignment process. We further employ a NCE loss \cite{nce} to optimize this alignment, defined as follows:
{\small
\begin{equation}
\begin{split}
\mathcal{L}_{\text{TQ}} = &- \frac{1}{|P'|} \sum_{\mathbf{p'} \in P'} 
\Bigg[ 
\log \sigma \left( \mathbf{p'}^\top \mathbf{t}^p \right) \\
&+ \frac{1}{|T^n|} \sum_{\mathbf{t^n} \in T^n} 
\log \left( 1 - \sigma \left( \mathbf{p'}^\top \mathbf{t}^n \right) \right) 
\Bigg],
\end{split}
\label{eq:q_t_loss}
\end{equation}}where \( \mathbf{t}^p \) and \( \mathbf{t}^n \) denote the representations of the positive and negative tags, respectively. The positive tag is labeled by the Stack Overflow platform, while negative tags are randomly sampled from the entire tag set. \( |P'| \) and \( |T^n| \) represent the total number of questions and the number of negative tags per question, respectively. 

During the offline training, we use the Alternating Least Square (ALS) \cite{als} technique to train the model, i.e., repeatedly optimize one of $\mathcal{L}_{\text{QQ}}$ and $\mathcal{L}_{\text{TQ}}$. \\
\subsection{Noise Tolerance Modeling} 
Prior work \cite{gbs2, gbs1, nf2, nf3, askcq2} on clarifying questions make a strong assumption: users always know the correct answer to a clarifying question. In the context of cQA, this implies that users are assumed to be 100\% certain about a query’s relevance to a tag. However, in practice, users may provide noisy answers due to limited domain knowledge or simple operational errors. To address this issue, after the two-stage offline training steps, we introduce noise tolerance modeling to mitigate the impact of noisy feedback during the clarification process. This modeling aims to evaluate the semantic similarity between a question \( p \) and an associated tag \( t \), using fixed representations obtained from earlier training. We optimize the model using binary cross-entropy loss, defined as follows: 
{\small
\begin{equation}
\begin{split}
\mathcal{L}_{\text{NR}} = 
&- \frac{1}{|P|} \sum_{\mathbf{p} \in P} 
\Bigg[
y \cdot \log \, \left( \Theta_n([\mathbf{p}; \mathbf{t}^{p}]) \right) \\
&+  \frac{1}{|T^n|}\sum_{\mathbf{t}^{n} \in T^n} 
(1 - y) \cdot \log \left( 1 - \left( \Theta_n([\mathbf{p}; \mathbf{t}^{n}]) \right) \right)
\Bigg],
\end{split}
\label{eq:bce-all}
\end{equation}}where $\mathbf{p}$ denotes the question embedding, and $\mathbf{t}^p$ and $\mathbf{t}^n$ denote the representations of the positive and negative tags, respectively. $\Theta_n$ represents the parameters of the noise tolerance module, which consists of two linear layers and maps the concatenated representation $[\mathbf{p}; \mathbf{t}]$ to a relevance score \( r \). \( |P| \) and \( |T^n| \) denote the total number of questions and the number of negative tags per question, respectively.

\subsection{Tag-enhanced Conversational Retrieval}
\textbf{Question Ranking with TeCQR.} After tag-enhanced two-stage offline training, we obtained enhanced representations of queries, questions, and tags. These representations are then fixed and utilized to perform conversational question retrieval. In this section, we introduce how to retrieve questions based on the trained embeddings and user feedback.

After \(l\) rounds of conversation (where \(l > 0\)), we obtain \(l\) rounds of clarifying questions and the corresponding user feedback, denoted as \(S_{(q^t, f)} = \{(q^t_1, f_1), (q^t_2, f_2), \dots, (q^t_l, f_l)\}\). The probability of relevance for each candidate question in the \(l\)-th iteration is ranked according to:
{\small
\begin{equation}
\label{eq:reranking}
\resizebox{0.46\textwidth}{!}{$
P(\mathbf{c} \mid C, \mathbf{Q}, S_{(q^t, f)}) = 
\frac{
    \exp\left[ \mathbf{c} \cdot \left( W_Q \cdot \mathbf{Q} + \sum_{(q^t, f) \in S_{(q^t, f)}} W_t \cdot \mathbf{e}_{(q^t, f)} \right) \right]
}{
    \sum\limits_{\mathbf{c}' \in C} 
    \exp\left[ \mathbf{c}' \cdot \left( W_Q \cdot \mathbf{Q} + \sum_{(q^t, f) \in S_{(q^t, f)}} W_t \cdot \mathbf{e}'_{(q^t, f)} \right) \right]
},
$}
\end{equation}}where \(\mathbf{c}\) represents a candidate question from the candidate list \(C\), and \(\mathbf{Q}\) is the representation of initial query. \(W_Q\) and \(W_t\) are personalized weighting parameters. 
The term \(\mathbf{e_{(q^t, f)}}\) represents the embedding of a clarifying question $q^t$ with feedback \(f\), which is defined as:
{\small
\begin{equation}
\mathbf{e}_{q^tf} =
\begin{cases} 
\text{ask another}, & \Theta_n([\mathbf{Q}; \mathbf{t}^f]) \leq \alpha \\
\mathbf{t}^f, & \Theta_n([\mathbf{Q}; \mathbf{t}^f]) > \alpha \\
\end{cases}
\quad \text{where } f \in \{+, -\},
\label{eq:nf}
\end{equation}}where $\mathbf{t}^+$ and $\mathbf{t}^-$ denote the positive and negative tag representations. To guide the decision-making process, we introduce a confidence threshold $\alpha$. 
If the confidence score $\Theta_n([\mathbf{Q}; \mathbf{t}^f]) \leq \alpha$, we regard such a signal as ambiguous and select a suboptimal tag for asking another clarifying question. If $\Theta_n([\mathbf{Q}; \mathbf{t}^f]) > \alpha$, we assume the model's prediction aligns with the user's intent, and the feedback ($\mathbf{t}^+$ or $\mathbf{t}^-$) is accepted and integrated to the representation $\mathbf{e}_{q^{tf}}$. The same mechanism applies to both positive and negative feedback.

It is worth noting that, there may be no conversation between the system and the user (e.g., when the first iteration (i.e., \( l = 0 \)), the system has not yet asked any clarifying questions). In this case, $\mathbf{e}_{q^tf}$ is set to zero. In specific, our model can still retrieve a question \(q_u\) from the candidate question set \( C \) according to:
\begin{equation}
P(\mathbf{c} \mid C, \mathbf{Q}) = \frac{\exp(\mathbf{c} \cdot \mathbf{Q})}{\sum_{\mathbf{c'} \in C} \exp(\mathbf{c'} \cdot \mathbf{Q})}.
\label{eq:relevance_probability}
\end{equation} 
\textbf{Learning to Ask.} In previous sections, we described how to utilize queries and clarifying questions with user feedback to retrieve related questions. Here, we explain how to select the best-suited tag to generate a clarifying question to ask the user. The goal of this selection process is to identify high-reward tags that can maximize the information gain and minimize the number of question times. Inspired by \citet{gbs1}, and \citet{gbs2}, we employ the Generalized Binary Search (GBS) \cite{GBSself} strategy as our tag selection strategy. GBS is a greedy approach that learns to select a sequence of best-suited tags for clarifying questions. The selection process is formally defined as:
\begin{equation}
t_l = \arg\min_t \left| \sum_{c \in C} \left( 2 \mathds{1}\{t^{c} = 1\} - 1 \right) \pi_l(c) \right|,
\label{eq:tag_selection}
\end{equation}
where \(t_l\) is the tag selected in the \(l\)-th round, and \(t^{c}\) indicates whether the question \(c\) is related to the tag \(t\). If the question \(c\) is related to the tag \(t\), then \(\mathds{1}\{t^{c} = 1\} = 1\); otherwise, if \(c\) is unrelated to \(t\), then \(\mathds{1}\{t^{c} = 1\} = 0\). \(\pi_l(c)\) represents a contribution score, defined as: 
\begin{equation}
\pi_l(c) = \frac{1}{\text{index}_{c} + 1},
\label{eq:contribution_score}
\end{equation}
where \(\text{index}_{c}\) is the rank of question \(c\) in the candidate questions list during the \(l\)-th round. Once a specific tag is selected as a clarifying question, the user provides feedback based on this question. The feedback is given as \textquotedbl{}yes\textquotedbl{} or \textquotedbl{}no\textquotedbl{}. Details of the user simulator are provided in the \textit{User Simulator}.
\section{Experiments}
\textbf{Dataset.} Due to the lack of conversational question retrieval datasets, we construct a new dataset named \textit{StackOverflow-Tag}, to advance research in conversational question retrieval. We begin by collecting queries from related questions on the SO platform. For each query, we employ the BM25 algorithm \cite{bm25} to retrieve 20 candidate questions from the corpus. We then use the official SO API to collect associated tags for each question, and discard any questions without tags. The resulting \textit{StackOverflow-Tag} dataset contains 54,786 questions and 2,278 unique tags. We divide the data into a training set with 13,772 queries and a test set with 1,482 queries, providing a realistic dataset for evaluating conversational question retrieval methods.\\
\textbf{Evaluation Metrics.} We use Recall@k (k=1, 3, 5), and NDCG@k (k=3, 5, 10), Mean Average Precision, Mean Reciprocal Rank as metrics to evaluate the retrieval performance, following prior work such as \cite{RCNN, rmso, tagco}. \\
\textbf{Implementation Details.}
All experiments are conducted on an NVIDIA RTX A6000 GPU. We use All-MiniLM as the encoder backbone and train with SGD (initial learning rate 0.1, decayed to 0). The embedding size is set to 384, matching the pre-trained model. The threshold $\alpha$ for the noise tolerance module is set to 0.5. Baseline results are reported using their optimal hyperparameters. Additional parameter sensitivity studies on the number of clarifying questions, batch size, and negative samples, as well as detailed software and hardware specifications, are provided in the \textit{Appendix}. \\
\textbf{Comparison Methods.} 
\label{sec:baselines}
We compare TeCQR with several baselines, including \textbf{BM25} \cite{bm25}, \textbf{GRU} \cite{gru}, \textbf{RCNN} \cite{RCNN}, \textbf{RMSO} \cite{rmso}, \textbf{Sentence-BERT} \cite{sentencebert, usesenbert}, \textbf{TS-QR} \cite{questionstag}, \textbf{All-MiniLM} \cite{allminilm}, \textbf{Query-Tag}, and \textbf{TEcotag} \cite{tagco}. Detailed descriptions of all compared methods are available in the \textit{Appendix}.
\begin{table*}[t]
\centering
\begin{tabular*}{\textwidth}{@{\extracolsep{\fill}}lccccccccc}
\toprule
\textbf{Method} & \textbf{R@1} & \textbf{R@3} & \textbf{R@5} & \textbf{NDCG@3} & \textbf{NDCG@5} & \textbf{NDCG@10} & \textbf{MAP} & \textbf{MRR} \\ 
\midrule
BM25            & 0.258        & 0.318        & 0.387        & 0.280          & 0.308           & 0.352            & 0.321        & 0.324        \\
GRU             & 0.365        & 0.435        & 0.504        & 0.401          & 0.430           & 0.464            & 0.440        & 0.441        \\
RCNN            & 0.386        & 0.465        & 0.534        & 0.429          & 0.457           & 0.490            & 0.466        & 0.468        \\
RMSO            & 0.412        & 0.502        & 0.565        & 0.462          & 0.497           & 0.522            & 0.505        & 0.507        \\ 
\midrule
Sentence-BERT   & 0.228        & 0.405        & 0.521        & 0.332          & 0.379           & 0.443            & 0.378        & 0.380        \\
TS-QR           & 0.382        & 0.463        & 0.538        & 0.422          & 0.455           & 0.494            & 0.466        & 0.469 \\       
All-MiniLM      & 0.443        & 0.673        & 0.798        & 0.579          & 0.631           & 0.687            & 0.599        & 0.604        \\
Query-Tag       & 0.432        & 0.661        & 0.789        & 0.563          & 0.617           & 0.670            & 0.584        & 0.588        \\
TEcotag         & \underline{0.448} & \underline{0.675} & \underline{0.801} & 0.563 & 0.628           & 0.690            & \underline{0.605} & \underline{0.608} \\ 
\midrule
\textbf{TeCQR (0 CQ)}        & 0.447        & 0.671        & 0.792        & \underline{0.580} & \underline{0.632} & \underline{0.692} & 0.603        & 0.607        \\ 
\midrule
\textbf{$\text{TeCQR}_\text{random}$ (5 CQs)} & 0.446 & 0.675 & 0.793 & \underline{0.580} & 0.629 & 0.687 & 0.600 & 0.604 \\
\textbf{TeCQR (5 CQs)}       & \textbf{0.498$^{*}$} & \textbf{0.717$^{*}$} & \textbf{0.833$^{*}$} & \textbf{0.629$^{*}$} & \textbf{0.677$^{*}$} & \textbf{0.720$^{*}$} & \textbf{0.644$^{*}$} & \textbf{0.649$^{*}$} \\ 
\bottomrule
\end{tabular*}
\caption{TeCQR Performance of Question Retrieval on the \textit{StackOverflow-Tag} dataset. The symbol ``$^{*}$'' refers to a significant improvement compared to the TEcotag baseline at the \( p < 0.05 \) level using the two-tailed pairwise t-test.} 
\label{tab:main_result}
\end{table*} \\
\textbf{User Simulator.}
\label{sec:User Simulator} Following prior work \cite{gbs2, gbs1, nf2, nf3}, we assume users can judge whether a query is related to a given tag. Based on this assumption, we design it as follows: if the tag in the clarifying question is related to the questions, the simulator responds with \textquotedbl{}yes\textquotedbl{}; otherwise, it responds with \textquotedbl{}no\textquotedbl{}. 
To better reflect real-world noise, in section \textit{Effects of Noise Tolerance Modeling}, we conducted experiments under the assumption that the user simulator has a certain probability of making errors, and we provide a detailed analysis of the impact of noisy user feedback and our noise tolerance.
\begin{figure}[!t]
    \centering
    \begin{subfigure}[b]{0.48\columnwidth}
        \centering
        \includegraphics[width=\textwidth]{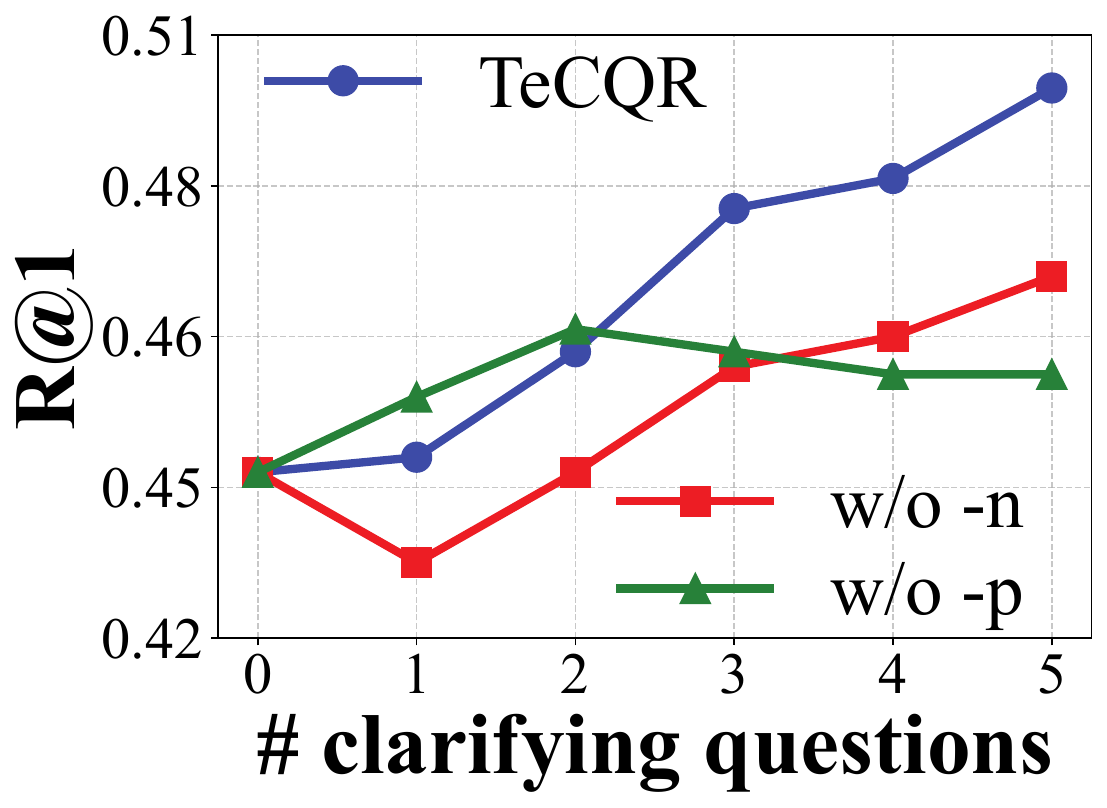}
        \caption{R@1}
        \label{fig:r1}
    \end{subfigure}
    \hfill
    \begin{subfigure}[b]{0.48\columnwidth}
        \centering
        \includegraphics[width=\textwidth]{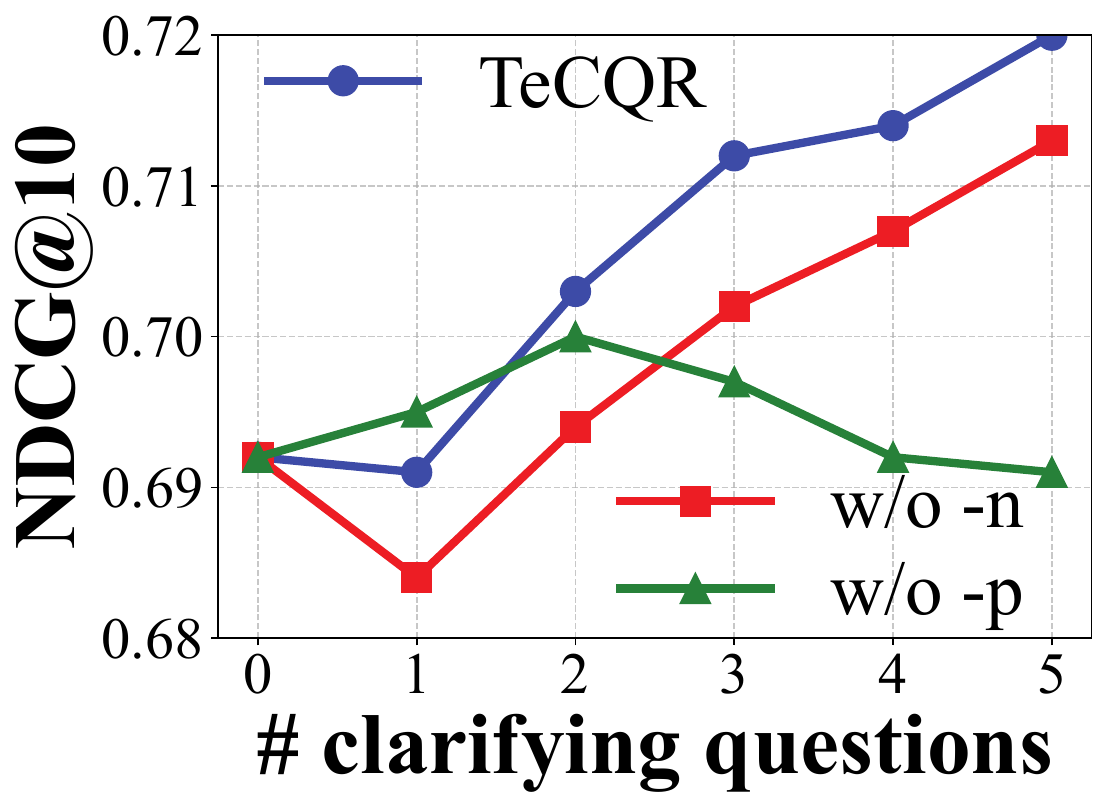}
        \caption{NDCG@10}
        \label{fig:ndcg}
    \end{subfigure}
    \begin{subfigure}[b]{0.48\columnwidth}
        \centering
        \includegraphics[width=\textwidth]{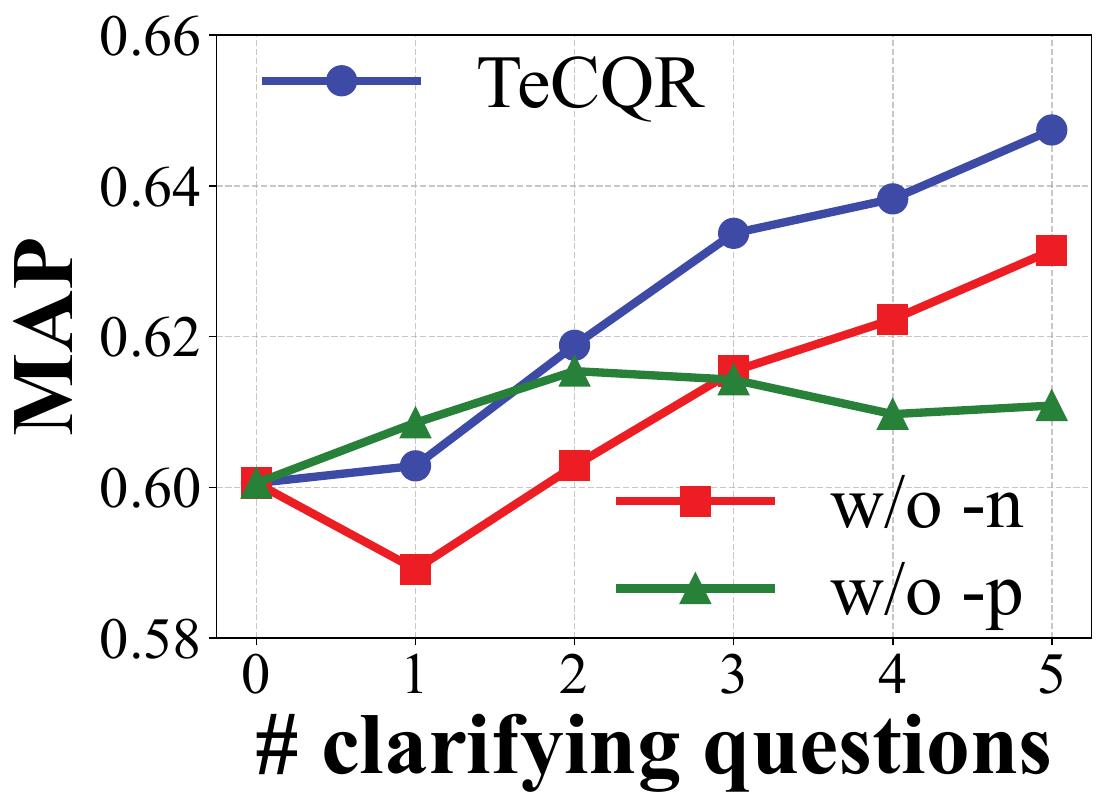}
        \caption{MAP}
        \label{fig:map}
    \end{subfigure}
    \hfill
    \begin{subfigure}[b]{0.48\columnwidth}
        \centering
        \includegraphics[width=\textwidth]{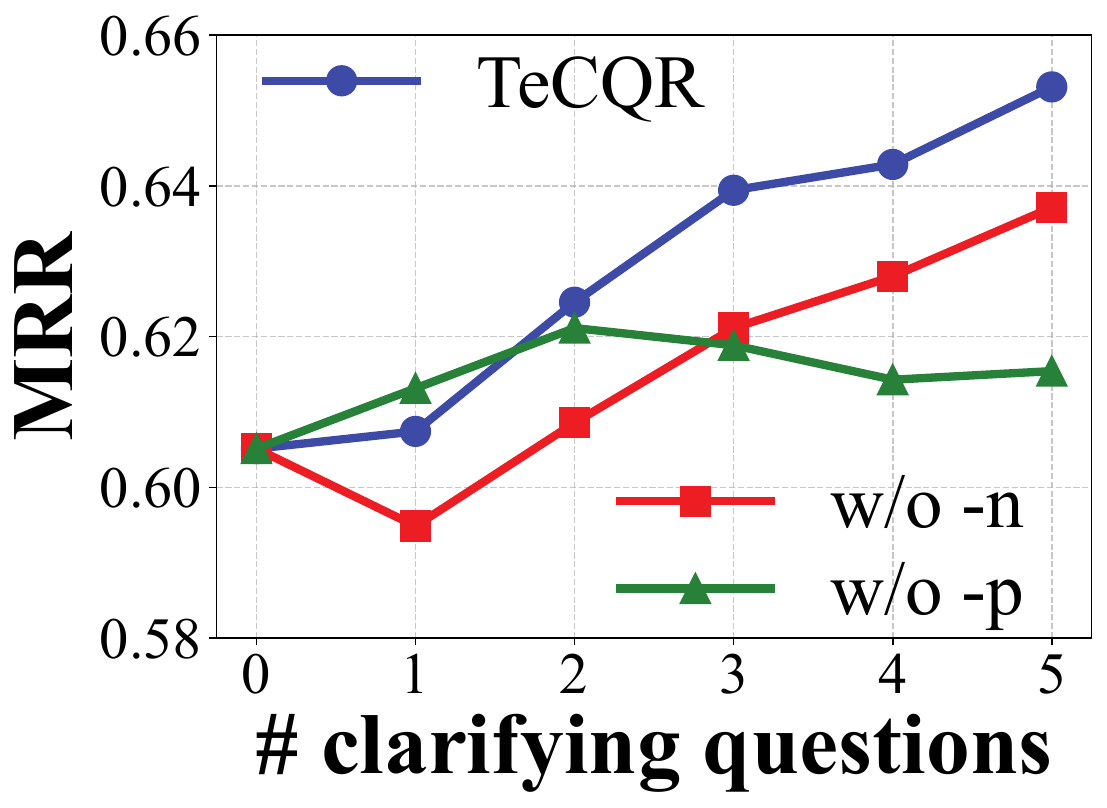}
        \caption{MRR}
        \label{fig:mrr}
    \end{subfigure}
    \caption{Effects of positive and negative user feedback.}
    \label{fig:p_n}
\end{figure}

\begin{figure}[!t]
    \centering
    \begin{subfigure}[b]{0.48\columnwidth}
        \centering
        \includegraphics[width=\textwidth]{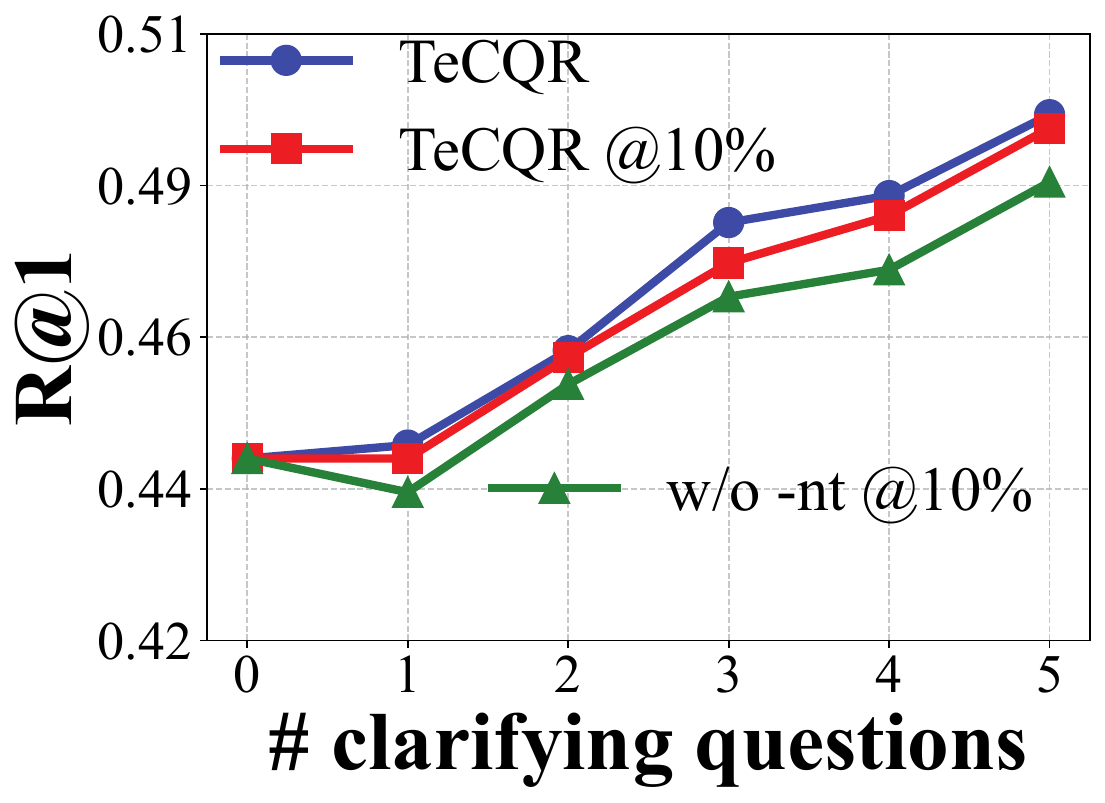}
        \caption{10\% noisy user feedback}
        \label{fig:5n}
    \end{subfigure}
    \hfill
    \begin{subfigure}[b]{0.48\columnwidth}
        \centering
        \includegraphics[width=\textwidth]{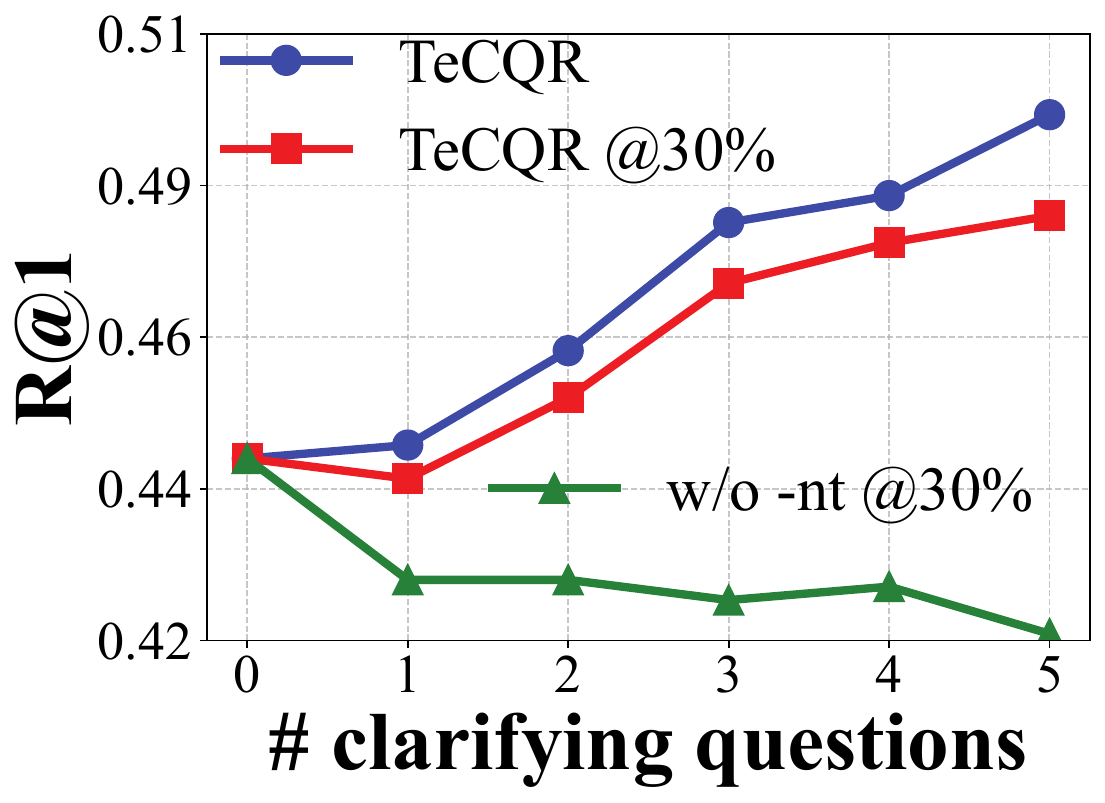}
        \caption{30\% noisy user feedback}
        \label{fig:15n}
    \end{subfigure}
    
    \begin{subfigure}[b]{0.48\columnwidth}
        \centering
        \includegraphics[width=\textwidth]{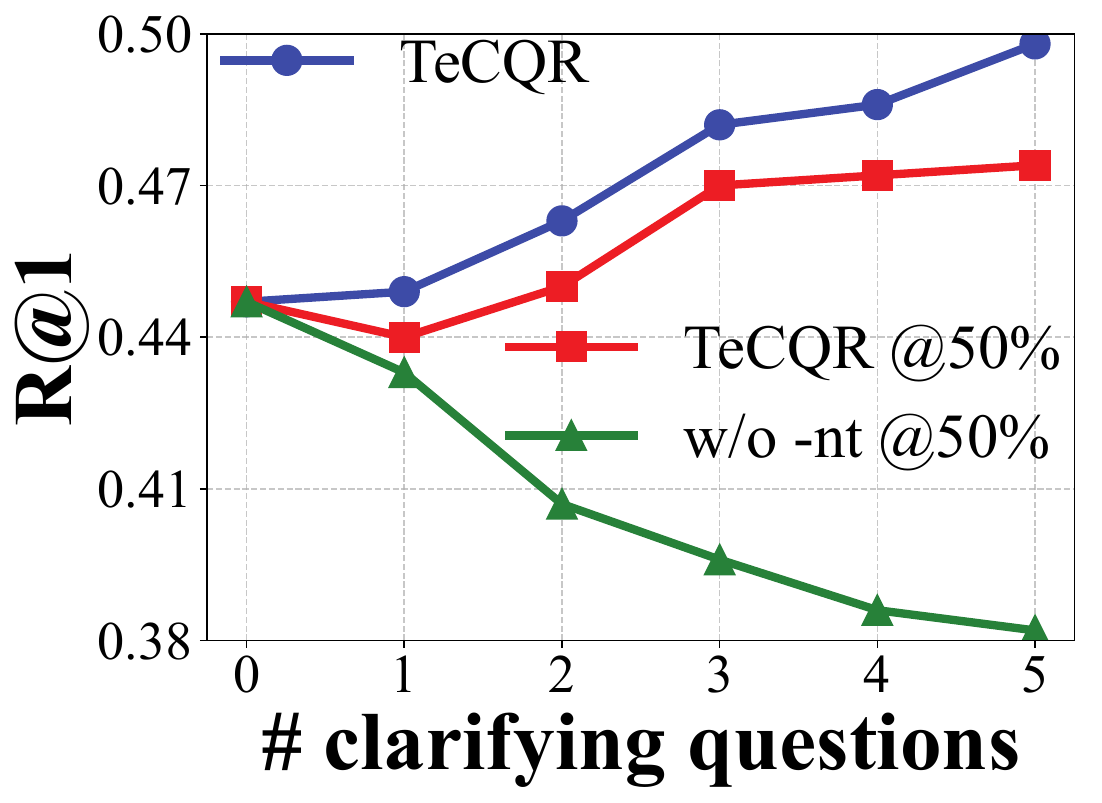}
        \caption{50\% noisy user feedback}
        \label{fig:25n}
    \end{subfigure}
    \hfill
    \begin{subfigure}[b]{0.48\columnwidth}
        \centering
        \includegraphics[width=\textwidth]{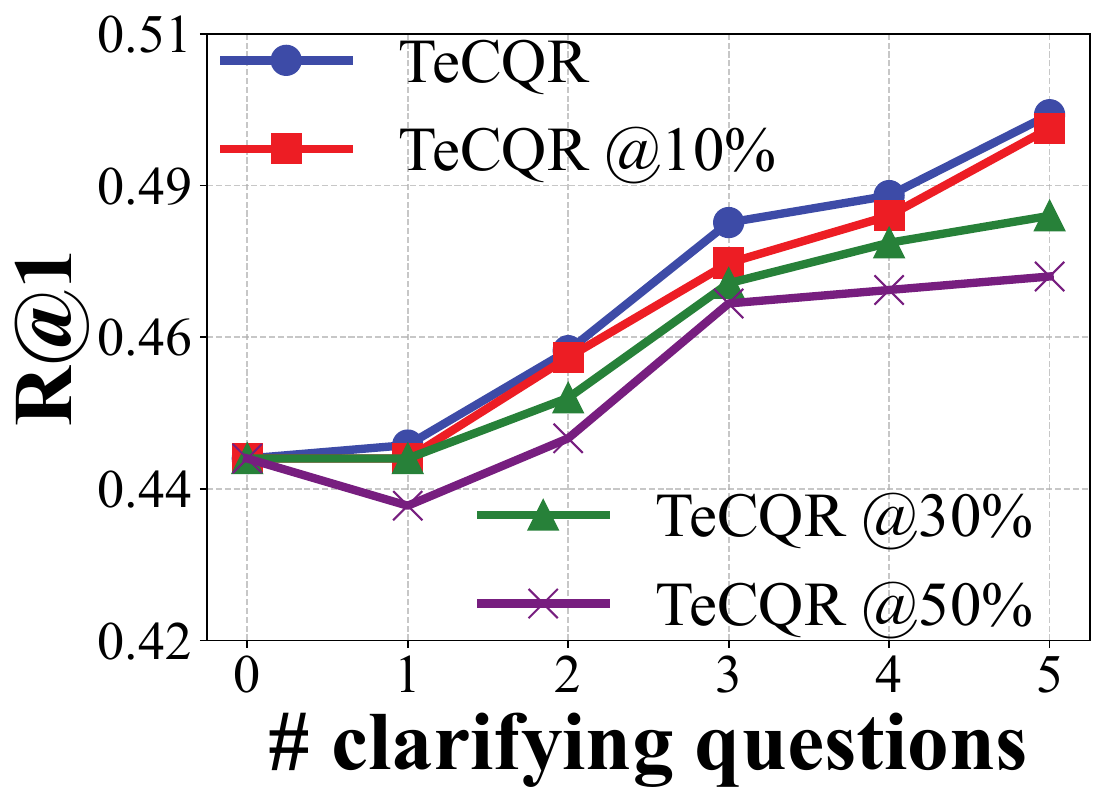}
        {\small
        \caption{various noisy user feedback}}
        \label{fig:alln}
    \end{subfigure}
    \caption{Effects of noisy user feedback.}
    \label{fig:noise}
\end{figure}
\subsection{Overall Performance Comparison}
As shown in Table \ref{tab:main_result}, BM25 performs the worst due to its lack of semantic modeling. GRU and RCNN improve results through deeper representations, while RMSO further enhances performance by fusing query and candidate embeddings. Although Sentence-BERT benefits from pre-training, it yields limited gains and even underperforms BM25 on R@1. TS-QR improves results by incorporating tag information via DeepWalk. All-MiniLM, trained on large-scale semantic matching data, achieves the best baseline results on the \textit{StackOverflow-Tag} dataset. Naive tag integration methods, such as Query-Tag (concatenation) and TEcotag (node2vec), show limited or even negative gains, highlighting the challenge of leveraging tag information.

Our $\text{TeCQR}$ with 5 CQs significantly outperforms all baselines, achieving MAP and MRR improvements of 6.4\% and 6.7\% over TEcotag. R@1 increases by 10\%, NDCG@10 by 4\%, with further gains in R@3 (+6.2\%), R@5 (+3.9\%), NDCG@3 (+11.7\%), and NDCG@5 (+7.8\%), demonstrating strong top-k performance. Even without clarification, $\text{TeCQR}$ (0 CQs) shows slight gains due to tag-enhanced two-stage offline training. Building on this, our core contribution lies in introducing conversation into question retrieval, where tag-enhanced two-stage offline training helps the system better handle short and ambiguous queries. A variant with random tag selection $\text{TeCQR}_\text{random}$ shows degraded performance as CQs increase, highlighting the importance of effective tag selection via GBS.
\subsection{Effects of Positive and Negative User Feedback} 
\label{sec:RQ2}
After receiving tag-based CQs, users give positive or negative feedback to indicate tag relevance. We evaluate two variants: (1) \textbf{TeCQR w/o -n}, which incorporates tags with positive feedback; and (2) \textbf{TeCQR w/o -p}, which incorporates tags with negative feedback. As shown in Figure~\ref{fig:p_n}, TeCQR w/o -n shows an initial drop but steady long-term gains, while TeCQR w/o -p yields slight early improvement with diminishing returns. By combining both feedback and adaptively weighting tag contributions, TeCQR achieves the best overall performance, demonstrating the effectiveness of modeling both positive and negative feedback.
\begin{figure}[!t]
    \centering
    \begin{subfigure}[b]{0.48\columnwidth}
        \centering
        \includegraphics[width=\textwidth]{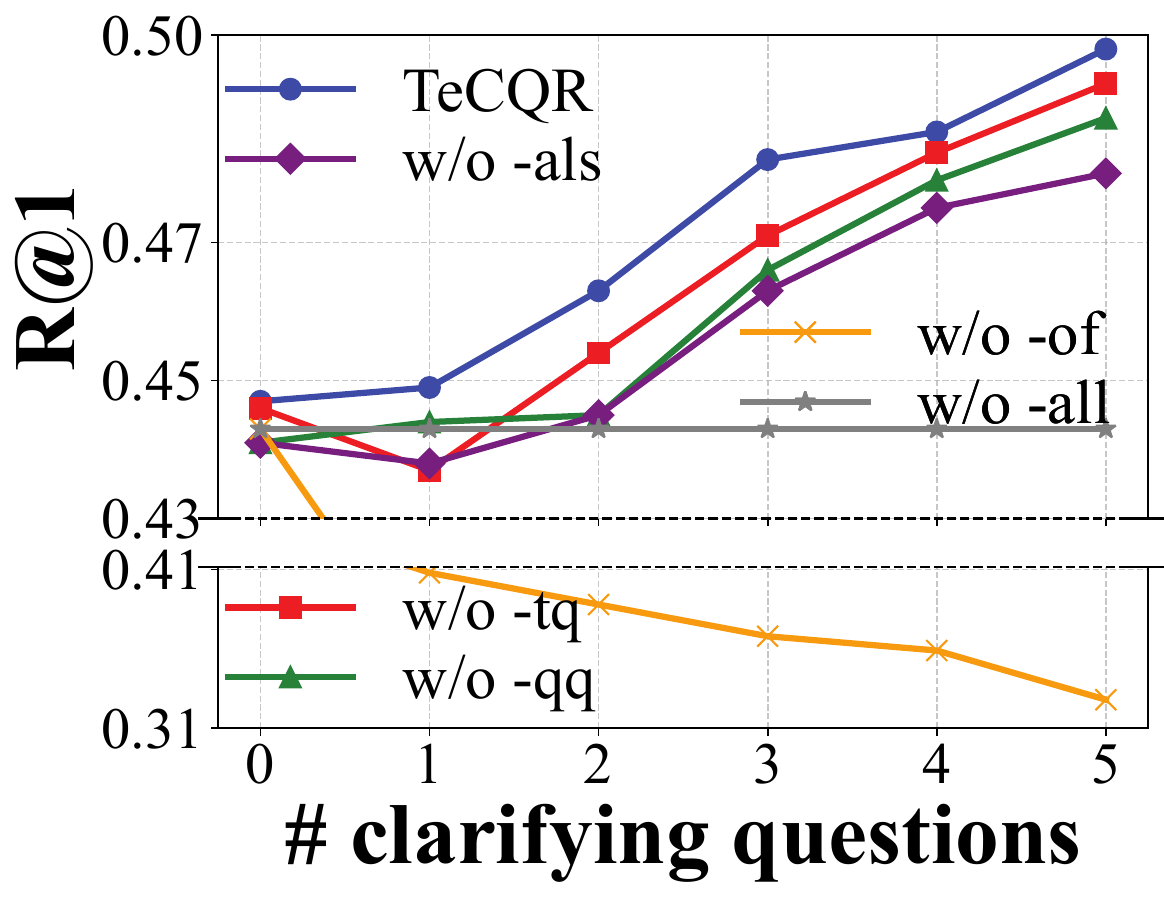}
        \caption{R@1}
        \label{fig:abr1}
    \end{subfigure}
    \hfill
    \begin{subfigure}[b]{0.48\columnwidth}
        \centering
        \includegraphics[width=\textwidth]{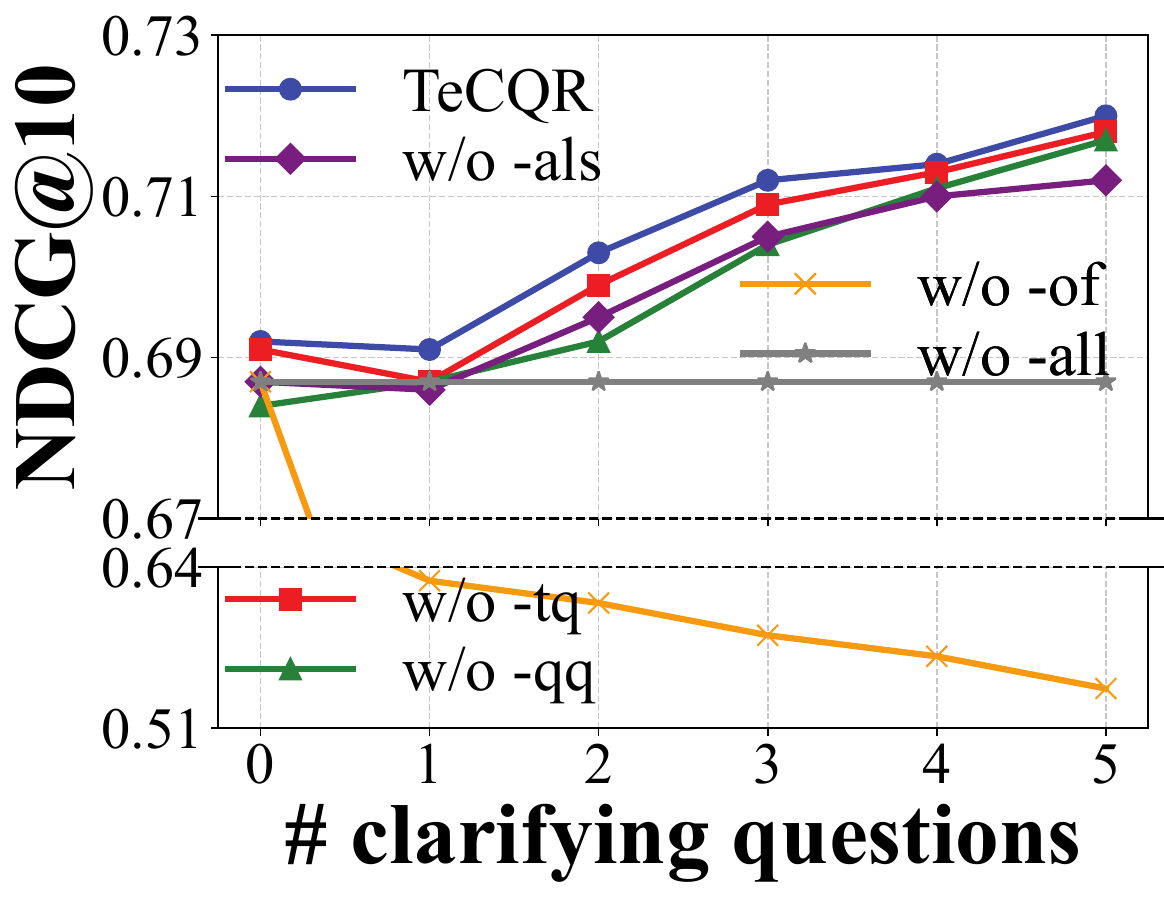}
        \caption{NDCG@10}
        \label{fig:abndcg}
    \end{subfigure}
    \begin{subfigure}[b]{0.48\columnwidth}
        \centering
        \includegraphics[width=\textwidth]{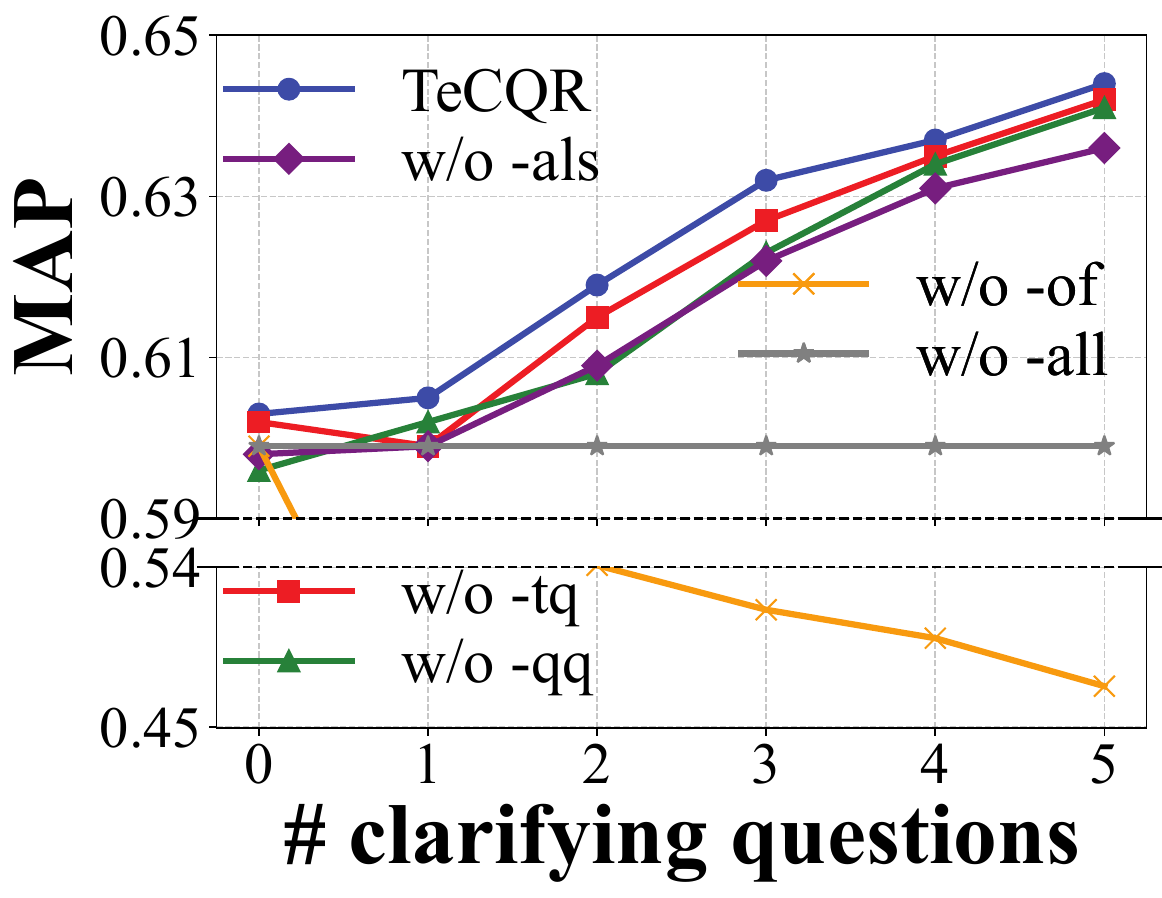}
        \caption{MAP}
        \label{fig:abmap}
    \end{subfigure}
    \hfill
    \begin{subfigure}[b]{0.48\columnwidth}
        \centering
        \includegraphics[width=\textwidth]{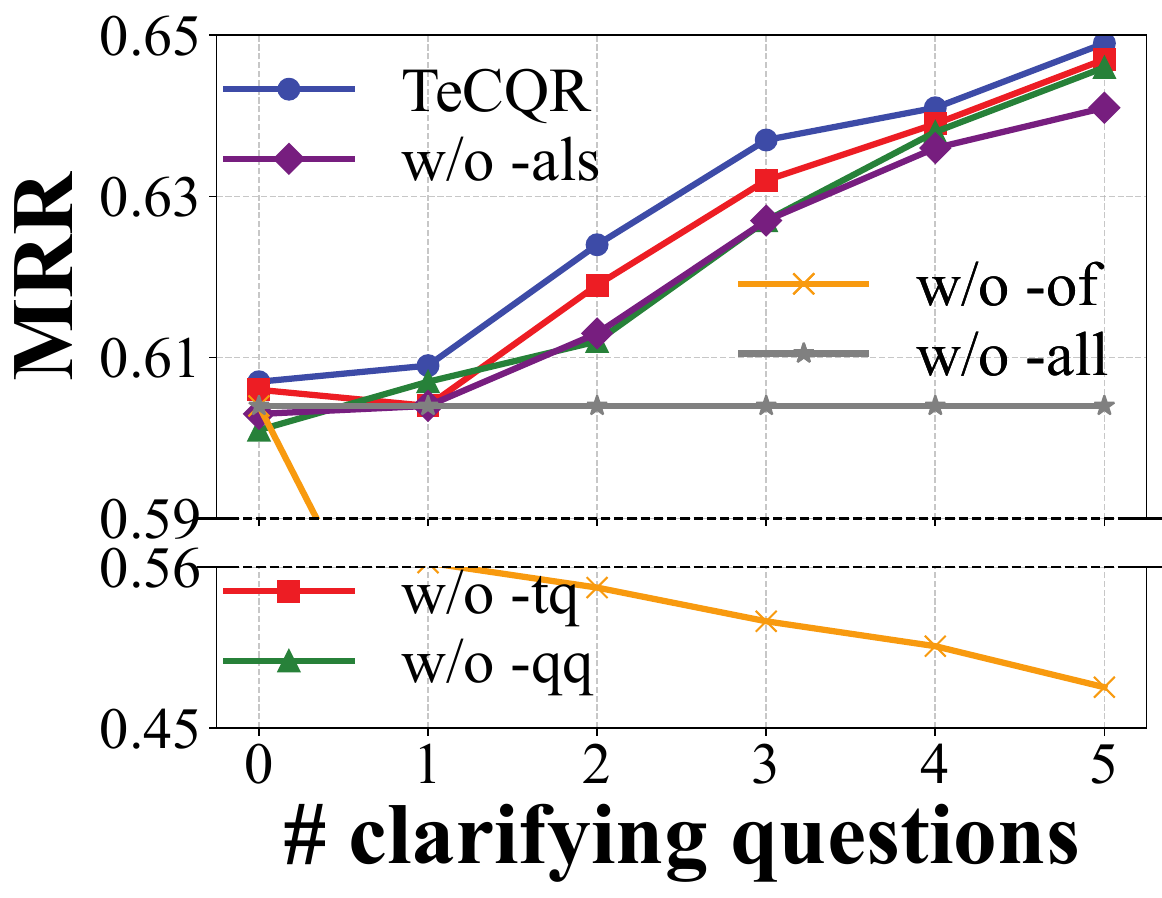}
        \caption{MRR}
        \label{fig:abmrr}
    \end{subfigure}
    \caption{Ablation Study.}
    \label{fig:ablation}
\end{figure}
\subsection{Effects of Noise Tolerance Modeling} 
\label{sec:noisy user feedback}
Following prior work~\cite{gbs2, gbs1, nf2, nf3}, we assume users can judge query–tag relevance, but real users may provide noisy feedback. To address it, we introduce noisy feedback with error rates from 10\% to 50\%, in steps of 20\%. We evaluate both the TeCQR and a variant without the noise tolerance module \textbf{TeCQR w/o -nt} under different noise condition. Results are shown in Figure~\ref{fig:noise}, where $k$ in \textbf{TeCQR @k\%} and \textbf{TeCQR w/o -nt @k\%} represents the error rate. As shown in Figure~\ref{fig:noise}, performance degrades as noise increases, which is expected. Nevertheless, TeCQR consistently benefits from CQs across all noise levels, while removing noise tolerance leads to significant drops. When noise exceeds 30\%, CQs even harm performance without noise tolerance, demonstrating both the vulnerability to noisy feedback and the effectiveness of our noise tolerance module.
\subsection{Contribution of Key Components} 
We conduct an ablation study to assess the contribution of each component in TeCQR (Figure~\ref{fig:ablation}). We compare: (1) \textbf{TeCQR w/o -tq}, removing tag-question training; (2) \textbf{TeCQR w/o -qq}, removing query-question training; (3) \textbf{TeCQR w/o -of}, directly applying All-MiniLM in the conversational retrieval phase without our tag-enhanced two-stage offline training; (4) \textbf{TeCQR w/o -als}, replacing ALS strategy with a standard joint loss; and (5) \textbf{TeCQR w/o -all}, removing all proposed modules and using All-MiniLM for static retrieval. All components contribute to performance. TeCQR w/o -of performs worst, confirming the value of two-stage offline training; removing either training objective (w/o -tq or w/o -qq) degrades performance, demonstrating their complementary benefits; TeCQR w/o -als is competitive but underperforms the ALS strategy; and TeCQR w/o -all remains a strong static baseline. Overall, TeCQR achieves the best results, validating our overall design.

\begin{figure}[t]
  \centering
  \includegraphics[width=0.99\linewidth]{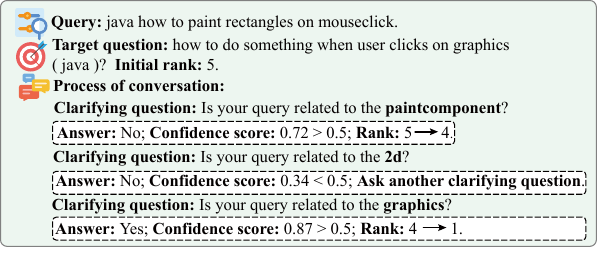} 
  \caption{Case Study.}
  \label{fig:casestudy}
\end{figure}
\subsection{Case Study} 
We present a case from the \textit{StackOverflow-Tag} dataset to illustrate TeCQR's conversational retrieval process. Initially, the target question was ranked 5. After the first clarifying question and a negative user response, its rank improved to 4. In the next round, the model's prediction from the noise tolerance model deviated from the user feedback. As a result, another clarifying question was asked, which received positive feedback. Ultimately, the rank of the target question improved to 1, demonstrating TeCQR's effectiveness in refining retrieval through conversation.

\section{Conclusion}
In this paper, we propose TeCQR, a novel model for conversational question retrieval in cQA platforms. It tackles the challenges of short, ambiguous queries and unreliable external information by introducing a novel multi-turn conversation into the question retrieval, allowing users to gradually clarify their intent. The TeCQR model mainly consists of three core modules: (1) a tag-enhanced two-stage offline training framework designed to improve the representations for more effective conversational retrieval, including query–question training via conversation simulation and tag–question training via contrastive learning; (2) a noise tolerance module that mitigates the impact of noisy user feedback; and (3) a tag-enhanced conversational retrieval module, which selects optimal tags to generate CQs and iteratively refines the query representation based on user feedback, enabling accurate related question retrieval.

\section{Acknowledgments}
This research was supported by the National Natural Science Foundation of China (62402093), the Sichuan Science and Technology Program (2025ZNSFSC0479), and the Fundamental Research Funds for the Central Universities (JBK202511020). This work was also supported in part by the National Natural Science Foundation of China under grants U20B2063 and 62220106008, and the Sichuan Science and Technology Program under Grant 2024NSFTD0034.
\small
\bigskip
\bibliography{ao}

\end{document}